# ROIsGAN: A Region Guided Generative Adversarial Framework for Murine Hippocampal Subregion Segmentation

Sayed Mehedi Azim, Brian Corbett, and Iman Dehzangi

*Abstract—* **The hippocampus, a critical brain structure involved in memory processing and various neurodegenerative and psychiatric disorders, comprises three key subregions: the dentate gyrus (DG), Cornu Ammonis 1 (CA1), and Cornu Ammonis 3 (CA3). Accurate segmentation of these subregions from histological tissue images is essential for advancing our understanding of disease mechanisms, developmental dynamics, and therapeutic interventions. However, no existing methods address the automated segmentation of hippocampal subregions from tissue images, particularly from immunohistochemistry (IHC) images. To bridge this gap, we introduce a novel set of four comprehensive murine hippocampal IHC datasets featuring distinct staining modalities: cFos, NeuN, and multiplexed stains combining cFos, NeuN, and either ΔFosB or GAD67, capturing structural, neuronal activity, and plasticity associated information. Additionally, we propose ROIsGAN, a region-guided U-Net-based generative adversarial network tailored for hippocampal subregion segmentation. By leveraging adversarial learning, ROIsGAN enhances boundary delineation and structural detail refinement through a novel region guided discriminator loss combining Dice and binary cross-entropy loss. Evaluated across DG, CA1, and CA3 subregions, ROIsGAN consistently outperforms conventional segmentation models, achieving performance gains ranging from 1-10% in Dice score and up to 11% in Intersection over Union (IoU), particularly under challenging staining conditions. Our work establishes foundational datasets and methods for automated hippocampal segmentation, enabling scalable, high-precision analysis of tissue images in neuroscience research. Our generated datasets, proposed model as a standalone tool, and its corresponding source code are publicly available at: https://github.com/MehediAzim/ROIsGAN**

## I. INTRODUCTION

The hippocampus is one of the most extensively studied areas in the brain because of its significant functional role in memory processing, its remarkable plasticity, and its involvement in epilepsy, neurodegenerative diseases, and psychiatric disorders [1-3]. Damage to the hippocampus can result in the loss of the ability to convert short-term memory to long-term memory, leading to cognitive dysfunction and mood disorders [4]. The hippocampus consists of three distinct regions: the DG, CA1, and CA3, all of which are crucial for information processing. For instance, adult neurogenesis takes place in the DG, the CA3 subregion is involved in the spatiotemporal processing of memory, and lesions to CA3 result in impaired spatiotemporal processing [4]. Furthermore, dorsal CA1 mediates topological memory, while dorsal DG, dorsal CA3, and dorsal CA1 mediate metric memory. Lesions in these regions can result in significant cognitive deficits [5].

Accurate identification and delineation of these brain regions play a pivotal role in advancing our understanding of neurodegenerative diseases, monitoring developmental dynamics, and assessing the efficacy of therapeutic interventions [6]. Various cutting-edge imaging techniques, such as immunohistochemistry (IHC), multiplex imaging, flow cytometry, intravital microscopy, and mass spectrometry imaging, have been developed to analyze tissue images [7-9]. Among these, IHC images stand out for their ability to capture the intricate molecular composition of different brain regions, enabling detailed visualization of neuronal subpopulations and protein expression patterns [10]. However, the inherent complexity of fluorescent IHC images characterized by irregular morphologies, varying signal intensities, and low signal-to-noise ratio poses significant challenges for accurate manual or semi-automated segmentation. Manually segmenting these regions is also time-intensive and prone to inter-observer variability, underscoring the need for automated, robust, and scalable solutions to analyze large-scale neuroscientific imaging datasets efficiently.

Deep learning (DL) methodologies have achieved unprecedented success in image analysis tasks, allowing for remarkable advancements in the ever-expanding landscape of medical image analysis, particularly neuroimaging [11]. Several studies have addressed brain tissue image segmentation, intending to improve precision and efficiency. These studies broadly fall into five categories: manual, region-based, thresholding-based, clustering-based, and feature extraction/classification-based approaches [12]. However,

Sayed Mehedi Azim is with the Center for Computational and Integrative Biology, Rutgers University, Camden, NJ 18103, USA (e-mail: sayedmehedi.azim@rutgers.edu).

Brian Corbett is with the Center for Computational and Integrative Biology, and the Department of Biology, Rutgers University, Camden, NJ 18103, USA (e-mail: brian.corbett@rutgers.edu).

Iman Dehzangi is with the Center for Computational and Integrative Biology, and Department of Computer Science Rutgers University, Camden, NJ 18103, USA. He is also affiliated with Rutgers Cancer Institute, Rutgers University, New Brunswick, NJ 08901, USA (e-mail:i.dehzangi@rutgers.edu).

these methods can be time-consuming, subjective, and susceptible to inter-observer variability.

Recent advancements in generative adversarial networks (GANs) have demonstrated remarkable potential for tissue image segmentation by enhancing structural details and segmentation performance in noisy or complex images [13-16]. Conditional GANs (cGANs), for instance, have improved pixel-wise accuracy in organ and tumor segmentation, while U-Net based GAN hybrids have enhanced biomedical image segmentation by integrating structural learning with adversarial loss [17-19]. However, most existing GAN-based methods focus on cell-level segmentation or single-modality imaging and lack specificity for anatomically distinct regions in complex, multi-stained IHC images.

While several DL-based studies have focused on cell segmentation and quantification [20-22], only a limited number have emphasized on segmenting region of interests (ROIs) from images captured using different imaging modalities. For example, Schell et al. investigated six deep learning-based methods to segment hippocampal regions that were done in magnetic resonance imaging (MRI) [23]. Additionally, Takeuchi et al. developed an ML-based model called CAseg to segment hippocampal CA2 area with an F1 score of 0.8 from Nissl-counterstained images [24]. To the best of our knowledge, no study has automated the segmentation of hippocampal subregions DG, CA1, and CA3 from fluorescent IHC images. This limitation hampers the ability to perform large-scale quantitative analyses of hippocampal subregion-specific alterations in neurological disorders.

To address these gaps, we introduce a novel set of comprehensive murine hippocampal tissue image datasets, featuring four distinct staining profiles: cFos, NeuN, and multiplexed combining cFos, NeuN, and either ΔFosB or GAD67. Additionally, we propose ROIsGAN, a region-guided U-Net-based generative adversarial network designed for hippocampal subregion segmentation. By leveraging adversarial learning, ROIsGAN enhances boundary delineation and structural detail refinement, advancing precision and efficiency in neuroscientific tissue image analysis.

The key contributions of our work can be summarized as follows.
- We introduce the first-of-its-kind dataset for hippocampal region segmentation, comprising images of the DG, CA1, and CA3 subregions along with their corresponding segmentation mask annotations across distinct staining modalities: cFos, NeuN, and multiplexed (cFos, NeuN, either ΔFosB or GAD67).
- We propose a novel generative adversarial network, called ROIsGAN, for segmentation of these hippocampal subregions, achieving significant performance across all four datasets.
- We propose a region-guided hybrid loss function for the discriminator, designed to enforce structural alignment between the ground truth and generated masks. By incorporating Dice score into the discriminator loss, our method emphasizes spatial overlap, which is particularly beneficial for segmenting small and irregularly shaped ROIs like the hippocampus, ultimately enhancing the generator's ability to produce more accurate and anatomically consistent segmentations.

## II. METHODS

The following subsections detail the experimental methodology, including the development of a novel hippocampal IHC benchmark dataset, data acquisition protocols, image preprocessing techniques, network architecture design, training strategy, and evaluation metrics employed to develop and validate the proposed ROIsGAN framework.

### A. Data Acquisition and Preprocessing

#### i. Mice and tissue preparation

10 to 14-week-old male and female C57BL/6 mice were used in this study. Mice were kept on a 12-12 light-dark cycle with lights being turned on at 7:00 and lights off at 19:00. For tissue collection, mice were transcardially perfused with saline. Left hemibrains were postfixed in 4% paraformaldehyde (PFA) in phosphate buffered saline (PBS) for 48 hours, followed by 72 hours in cryoprotectant (30% sucrose in PBS). Cryoprotected brains were sectioned at 30 μm on a sliding microtome. All procedures were approved by the Rutgers University Institutional Animal Care and Use Committee.

#### ii. Immunohistochemistry

Murine tissue images were obtained from Rutgers University-Camden and the Joint Health Science Building following approved ethical guidelines. Immunohistochemistry was performed as previously described [25]. Primary antibodies used were monoclonal mouse anti-GAD-67 (1:1000, EMD Millipore, MAB5406), guinea pig anti-NeuN (1:3000, EMD Millipore, ABN90), rabbit anti-cFos (1:1000, Synaptic Systems, 226-008), mouse anti-cFos (1:500, Santa Cruz, sc-271243), and rabbit anti-FosB (1:100, Cell Signaling, #2263). Alexa Fluor™ goat anti-guinea pig 405 (Abcam, ab175678), goat anti-mouse 488 (Abcam, ab150113) and goat anti-rabbit 594 (Abcam, ab150080) were used as secondaries at a concentration of 1:200. Sections were blocked with 10% normal goat serum for one hour and incubated with primary antibodies overnight at 4°C. Sections were rinsed and incubated in secondary antibodies for 120 minutes. Sections were mounted and imaged on a Leica TCS SP8 confocal microscope at 10x. Cell counting was used to quantify c-Fos. NeuN was visualized in all sections. Either mouse-anti-cFos with rabbit-anti-FosB or rabbit-anti-cFos with mouse-anti-GAD67 were also used to visualize proteins. Sections from different mice using two different sets of antibodies were used to ensure that multiple protein expression patterns were used to train our algorithm.

### B. Benchmark Dataset

We collected data from 53 adult C57BL/6 mice, of which 27 were females and 26 were males. The dataset curated for this study consists of high-resolution IHC tissue images with manually annotated hippocampal regions annotated by three experts with domain knowledge, which serves as ground truth masks. Fiji ImageJ was used to annotate these images by creating an annotation mask for each region. Preprocessing steps, including normalization, histogram equalization, and stain normalization, were applied to reduce variations in staining protocols.

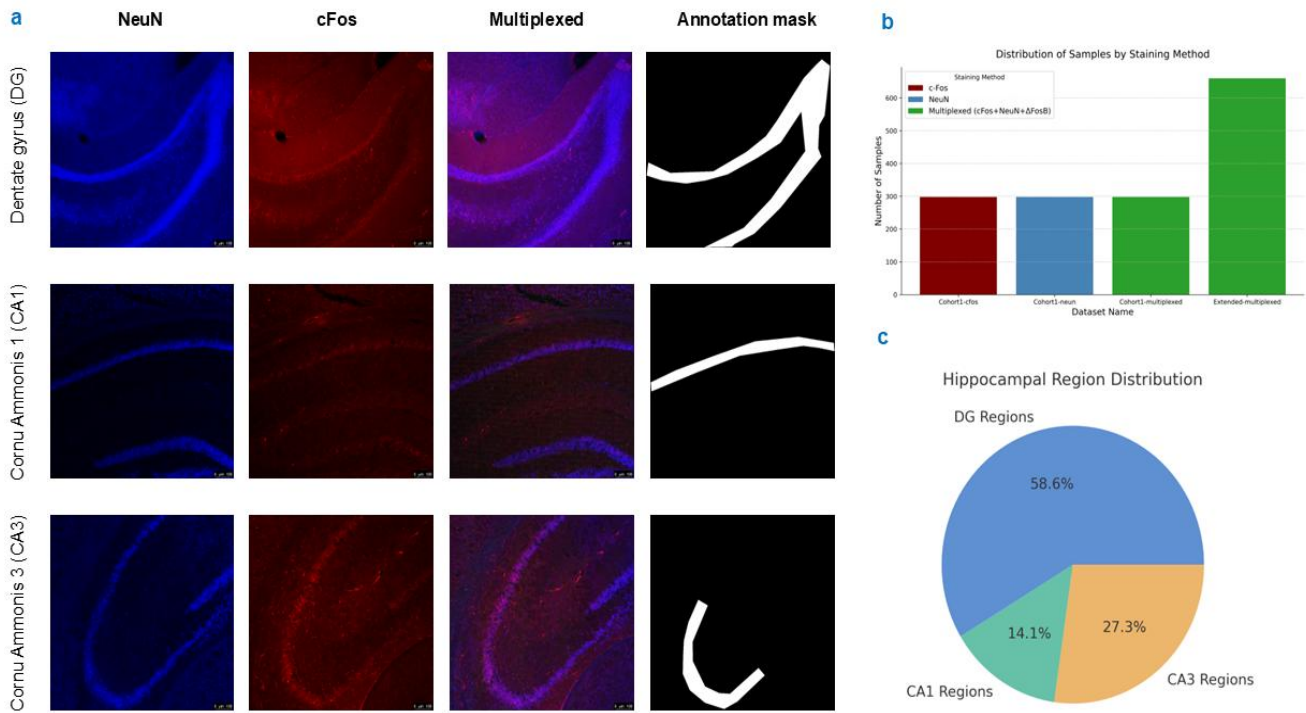

*Figure 1. Overview of presented dataset; a) Sample data for training ROI segmentation model, b) Data distribution in the proposed datasets, c) Distribution of hippocampal regions in the extended multiplexed dataset.*

For hippocampal region segmentation, four datasets were prepared, containing 1,260 unique samples with a total of 1,558 IHC images. The Cohort1-cFos dataset, containing cFos-stained images, includes 302 images. The Cohort1-NeuN dataset and the Cohort1-Multiplexed dataset contain 298 images each. To address an imbalance in CA1 annotations, we collected an additional multiplexed set, yielding an Extended-Multiplexed dataset with 660 images. The extended multiplexed dataset includes data collected using two staining protocols: one combining cFos, NeuN, and ΔFosB (as in the cohort 1 multiplexed dataset), and another combining cFos, NeuN, and GAD67, introducing additional anatomical variability.

Here, cFos serves as an activity marker, NeuN as a structural marker, and ΔFosB as a plasticity marker. cFos is a marker for neuronal activity that is sparsely expressed in neurons in the dentate gyrus granule cell layer and the neurons in the pyramidal cell layers of CA3 and CA1. NeuN is expressed in all neurons of the dentate gyrus granule cell layer and pyramidal cell layers of CA3 and CA1, allowing for clear visualization of neuronal nuclei/cell bodies throughout the hippocampus. ΔFosB is a marker for chronic neuronal activity and is also expressed in all neurons in the hippocampus, with intensities that have greater inter-mouse variability than those of NeuN.

GAD67 is primarily expressed in axons and dendrites in the hippocampus and not in the nucleus and soma. Hence, its expression pattern has a low degree of overlap with the other proteins that were visualized. Figure 1 provides a visual overview, with a) Example training data used for the hippocampal region segmentation model, b) Data distribution across the proposed datasets, and c) shows a pie chart illustrating the hippocampal region distribution in this extended dataset.

### C. Preprocessing

In the preprocessing step, all images were rescaled to the [0, 1] and resized from 1024×1024 to 256 × 256 pixels using bilinear interpolation via OpenCV to reduce computational cost. Z-score normalization was then applied using per-channel mean and standard deviation computed across the training set to improve model stability. Ground truth masks, provided in grayscale, were resized using nearest-neighbor interpolation to preserve binary boundaries and scaled to [0,1].

To improve model robustness to staining and orientation variability, data augmentation was applied during training using the Albumentations library. Augmentations included horizontal flipping (probability = 0.5), random rotation within ±10° (p = 0.3), and brightness/contrast adjustments (p = 0.2), followed by a forced resize to 256×256 to ensure dimensional consistency. Finally, the dataset was split into training (80%), validation (5%), and test (15%) subsets using a fixed random seed (42) to ensure reproducibility.

### D. Network Architecture

Our proposed ROIsGAN framework employs a generative adversarial network to segment hippocampal regions from murine IHC tissue images. A typical GAN consists of a generator and a discriminator. Initially, the generator is trained to create synthetic samples indistinguishable from the ground truth. On the other hand, the discriminator is trained to distinguish between real and synthetic samples. This framework can be represented as a two-player min-max game between the generator and discriminator, where each network iteratively improves through adversarial training [16]. In our work, this adversarial framework is adapted for segmentation by employing a U-Net-based generator that produces pixel-wise binary masks conditioned on input IHC images. Rather than using a discriminator as a binary classifier, we implement a U-Net-like architecture that outputs a spatial realism probability map. This design provides region-aware feedback to the generator, improving its ability to produce anatomically accurate segmentation masks. The discriminator is optimized using a hybrid loss that combines Dice and binary cross-entropy terms, further guiding the generator toward structure-preserving outputs. ROIsGAN is designed to generalize across varying staining protocols and consistently outperforms baseline models in segmentation performance across all modalities.

### E. The Generator

The generator in ROIsGAN adopts a four-level encoder-decoder U-Net architecture and is responsible for generating binary segmentation masks from input IHC images. As illustrated in Figure 2, the architecture consists of a symmetric encoder-decoder structure with continuous skip connections and a central bottleneck. These skip connections, combined with deep contextual encoding through the bottleneck, ensure that both global context and fine spatial detail are preserved [26], which is critical for accurately delineating small hippocampal subregions.

Each encoder block contains two consecutive sequences of 3×3 convolutional layers, followed by batch normalization and ReLU activation operation. Downsampling is performed using 2×2 max-pooling layers, and the number of feature channels doubles at each level, progressing from 3→64→128→256→512. The bottleneck block increases the feature dimensionality to 1024 channels, enabling the model to capture high-level semantic information at a resolution of 16×16.

The decoder mirrors the encoder structure with four upsampling stages, each beginning with a 2×2 transposed convolution to double the spatial resolution and halve the feature depth. Skip connections concatenate the corresponding encoder feature maps, preserving localization information. Each decoder stage includes a convolutional block identical in structure to the encoder block. Finally, a 1×1 convolution reduces the feature map to a single output channel, and a sigmoid activation is applied to generate the binary segmentation mask at 256×256 resolution. All convolutional layers are initialized using He (Kaiming) initialization to promote stable and efficient training [27].

### F. The discriminator

The discriminator in ROIsGAN is a two-level U-Net-like network designed to assess the realism of predicted segmentation masks while providing spatially dense adversarial feedback to the generator. It processes single-channel inputs of size 256×256 and produces a realism probability map enabling pixel-wise supervision. This design diverges from traditional GAN discriminators [28], integrating segmentation-specific feedback directly into the adversarial learning process.

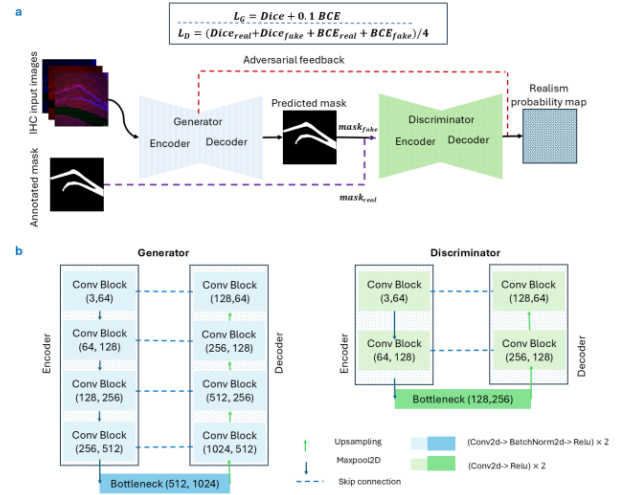

*Figure 2. Model architecture of the proposed ROIsGAN framework. (a) End-to-end pipeline illustrating ROIsGAN's workflow, in which the generator produces hippocampal segmentation masks from input IHC images, while the discriminator generates spatial realism probability maps to guide anatomically precise mask prediction. (b) Detailed architecture of the generator and discriminator networks, highlighting the U-Net-based encoder-decoder structure, skip connections, and the spatial realism assessment strategy of the discriminator.*

The encoder consists of two convolutional blocks, each featuring two consecutive sequences of 3×3 convolutional layer followed by ReLU activations. Feature dimensionality increases progressively from 1→64 and 64→128, with 2×2 max-pooling applied after each block. The bottleneck encodes the representation into 256 channels at a reduced resolution of 64×64 capturing multi-scale features efficiently.

The decoder upsamples using two transposed convolutional layers (2×2, stride=2), reducing channels depth from 256→128, and then 128→64. Skip connections concatenate encoder features with decoder inputs, allowing the model to retain spatial information. Each up-sampling stage is followed by a convolutional block similar to the block used in encoder, allowing the model to refine spatial details in reconstruction. A final 1×1 convolution followed by a sigmoid activation generates the realism map, where each pixel encodes the local probability of anatomical plausibility. Figure 2 depicts the complete architecture of our ROIsGAN framework.

### G. Loss Functions

The proposed ROIsGAN framework leverages a hybrid loss strategy to optimize segmentation accuracy and adversarial realism within a min-max game between the generator and discriminator. This approach prioritizes anatomical fidelity in segmenting ROIs while harnessing adversarial training to enhance boundary precision.

### i. Generator Loss

The generator loss ($L_G$) integrates Dice and adversarial binary cross-entropy (BCE) losses. The hybrid loss function is defined by

$$L_G = \lambda_{dice} L_{Dice}(y_g, y_p) + \lambda_{adv} L_{BCE}(D(y_p), 1) \quad (1)$$

Where and $\lambda_{dice} = 1.0$ and $\lambda_{adv} = 0.1$ was used. The Dice term,

$$L_{Dice} = 1 - \frac{2\sum y_g \cdot y_p}{\sum y_g + \sum y_p + \epsilon} \quad (\epsilon = 10^{-8}) \quad (2)$$

maximizes overlap between predictions $y_p$ and ground truth $y_g$. The adversarial BCE term $L_{BCE}(D(y_p), 1)$, penalizes deviations from realistic outputs, aligning with traditional GAN training [16].

### ii. Proposed Region-Guided Discriminator Loss

We introduce a novel blend of structural fidelity via Dice and realism assessment via BCE in the discriminator loss ($L_D$) which is defined by

$$L_D = \frac{(L_{Dice}(y_g, D(y_g)) + L_{Dice}(y_g, D(y_p)) + L_{BCE}(D(y_g), 1) + L_{BCE}(D(y_p), 0))}{4} \quad (3)$$

An additional gradient penalty was added to the $L_D$ for stabilizing the training. Unlike conventional discriminators that rely solely on BCE [16] or L1 losses [29], this formulation uniquely enforces structural alignment to improve the generator's ability to produce more accurate masks. Unlike L1 loss, which penalizes pixel-wise differences uniformly, Dice prioritizes overlap, better suiting small, irregular ROIs like the hippocampus, enhancing segmentation efficacy.

### H. Post-Processing

The proposed ROIsGAN framework integrated a post-processing step to refine segmentation masks using connected components analysis (CCA). Raw probability maps from the generator were binarized at τ=0.3, followed by 8-connectivity component analysis with area filtering, which enhanced region cohesion by accounting for diagonal pixel relationships. We excluded isolated components smaller than 300 pixels, a threshold determined through empirical analysis with pixel values from 5 to 400, to remove artifactual fragments typical in neuroimaging while retaining morphologically significant hippocampal structures. This approach aligns with artifact-reduction strategies in modern segmentation pipelines, where CCA-based post-processing improves structural coherence without retraining the model.

### I. Training protocol and experimental setting

The generator and discriminator were trained jointly in an adversarial setting using separate Adam optimizers. The generator used a learning rate of $5 \times 10^{-4}$, while the discriminator used a lower learning rate of $1 \times 10^{-5}$. The generator loss combined a segmentation objective (Dice loss) with an adversarial component (binary cross-entropy), weighted at a ratio of 1.0:0.1. The discriminator loss incorporated both Dice and binary cross-entropy terms for real and fake masks, along with a gradient penalty term (λ=1.0) to stabilize training. Training was conducted for up to 1000 epochs, with early stopping. The model achieving the highest validation Dice was retained for final evaluation. All experiments were implemented in PyTorch and conducted on an NVIDIA RTX A6000 48 GB GPU.

### J. Evaluation Metrics

In medical image segmentation, the primary objective is to quantify the similarity between predicted regions and ground truth annotations. The Dice Similarity Coefficient (Dice) and Intersection over Union (IoU) are well-established metrics for assessing segmentation performance, particularly in tasks requiring precise volumetric overlap, such as hippocampus tissue segmentation [30, 31].

To quantitatively assess the performance of our proposed region-guided adversarially trained network ROIsGAN for hippocampus tissue segmentation, we adopted a comprehensive set of metrics: Dice, IoU, Hausdorff Distance (HD), Precision, Recall, and Average Symmetric Surface Distance (ASSD). Each metric provides distinct insights into the model's segmentation accuracy, boundary delineation, and spatial consistency, enabling robust comparisons against baseline architectures.

The **Dice Similarity Coefficient (Dice)** measures the overlap between the predicted segmentation mask $y_p$ and the ground truth $y_g$ defined as:

$$Dice = \frac{2|y_p \cap y_g|}{|s_p| + |s_g|} \quad (4)$$

Dice emphasizes volumetric agreement, with higher values indicating better alignment. It remains a cornerstone metric in medical imaging due to its sensitivity to overlap [31].

The **Intersection over Union (IoU)**, also known as the Jaccard Index, quantifies the ratio of the intersection to the union of the predicted and ground truth regions:

$$IoU = \frac{|y_P \cap y_g|}{|y_P \cup y_g|} \quad (5)$$

IoU complements Dice by penalizing false positives and negatives more explicitly and has been widely adopted in computer vision and medical image analysis [30].

To evaluate boundary accuracy, we computed the **Hausdorff Distance (HD)**, which measures the maximum Euclidean distance between points on the predicted and ground truth boundaries:

$$HD = \max \left\{ \sup_{p \in \partial s_p} \inf_{g \in \partial s_g} d(p, g), \sup_{g \in \partial s_g} \inf_{p \in \partial s_p} d(g, p) \right\} \quad (6)$$

Reported in pixels, lower HD values reflect superior contour fidelity, critical for delineating the hippocampus's complex morphology. HD is particularly sensitive to boundary outliers thus plays an important role in assessing segmentation outliers [32].

**Precision** and **Recall** were included to assess the model's ability to balance true positive detections against over- and under-segmentation:

$$Precision = \frac{TP}{TP + FP} \quad (7)$$

$$Recall = \frac{TP}{TP + FN} \quad (8)$$

where TP, FP, and FN denote true positives, false positives, and false negatives, respectively. Both metrics, expressed as percentages, are foundational in evaluating classification performance in modern segmentation framework [33].

Finally, the **Average Symmetric Surface Distance (ASSD)** quantifies the average distance between the surfaces of the predicted and ground truth segmentations:

$$ASSD = \frac{1}{|\partial s_p| + |\partial s_p|} \left( \sum_{p \in \partial s_p} \inf_{g \in \partial s_g} d(p,g) + \sum_{g \in \partial s_g} \inf_{p \in \partial s_p} d(g,p) \right) \quad (9)$$

Reported in pixels, ASSD provides a symmetric measure of surface alignment, with lower values signifying closer agreement. This metric is particularly suited for assessing fine-grained boundary accuracy in medical applications [34].

These metrics collectively enable a thorough evaluation of segmentation quality, capturing both region-based overlap (Dice, IoU, Precision, Recall) and boundary-specific precision (HD, ASSD). Results were computed across the test dataset, with statistical significance assessed to validate the superiority of our approach over competing methods.

### K. Explainability Analysis

To interpret ROIsGAN's U-Net-based generator's segmentation decisions across all datasets, we employed Gradient-weighted Class Activation Mapping (Grad-CAM) [35]. Grad-CAM was selected for its ability to provide class specific spatial explanations, making it well suited for evaluating how the model localizes anatomical regions during segmentation. Grad-CAM was computed by extracting gradients of the predicted mask with respect to the decoder's final convolutional layer, which encodes high-level spatial features [35]. These gradients were globally averaged and weighted over the activation maps to produce heatmaps, which were normalized to the [0, 1] range using ReLU and min–max scaling.

Primary Grad-CAM maps were derived from predicted masks to capture model attention, while ground truth–driven maps were used for qualitative comparison. Visualizations included Grad-CAM heatmaps with predicted and true contours, overlays (30% transparency), and error maps highlighting true positives (green), false positives (red), and false negatives (blue). Heatmaps were rendered using the jet colormap for visual clarity.

### III. RESULTS

ROIsGAN was evaluated on four datasets introduced in this study: Cohort1-cFos, Cohort1-NeuN, Cohort1-Multiplexed, and the Extended-Multiplexed dataset. These datasets include a total of 1,558 stained images of hippocampus ROIs, encompassing diverse staining modalities for evaluating segmentation accuracy, and generalizability.

To benchmark the model's performance, we compared ROIsGAN against four widely used segmentation architectures namely, U-Net [26], U-Net++ [36], Attention U-Net [37], and ResU-Net++ [38] using a comprehensive evaluation framework based on the Dice coefficient, Intersection over Union (IoU), Hausdorff Distance (HD), Precision, Recall, and Average Symmetric Surface Distance (ASSD).

Each model was independently trained and evaluated on all four datasets, allowing for a robust assessment of generalization ability, boundary delineation, and performance stability under varying imaging conditions. ROIsGAN leverages a novel region-guided adversarial training strategy (as described in Methods), which synergistically enhances both segmentation accuracy and the boundary precision. The results, summarized in Tables 1–4, demonstrate that ROIsGAN consistently outperforms all baseline models across all modalities, particularly in datasets with complex or sparsely stained regions, such as cFos.

### A. Cohort 1-Multiplexed Dataset

In the Cohort 1 multiplexed dataset (Table 1), which combines cFos, NeuN, and ΔFosB stains, ROIsGAN achieves a mean Dice coefficient of 0.85, surpassing U-Net++ (0.83), the closest baseline. It also outperforms U-Net (0.81), Attention U-Net (0.82), and ResU-Net++ (0.82). This improvement reflects the enhanced segmentation performance enabled by the adversarial framework, particularly in complex multi-stained ROIs.

In addition to Dice, ROIsGAN excelled in boundary precision metrics, with a HD of 31.5 pixels and an ASSD of 4.2 pixels, compared to U-Net++'s 42.0 and 5.2, respectively. This reduction in boundary errors demonstrates the model's superior ability to capture fine-grained structural details and reduce over-segmentation artifacts. Furthermore, ROIsGAN achieved better IoU (0.75), Precision (0.84), and Recall (0.86) compared to other methods, underscoring its robustness in handling multi-stained datasets (Figure 3. e).

*Table 1. Performance comparison with existing state-of-the-art model on the cohort 1 Multiplexed image dataset*

| Model | Dice | IoU | HD | Precision | Recall | ASSD |
|---|---|---|---|---|---|---|
| Unet | 0.81 | 0.70 | 61.1 | 0.82 | 0.82 | 6.8 |
| Unet++ | 0.83 | 0.72 | 42.0 | 0.83 | 0.84 | 5.2 |
| AttentionUnet | 0.82 | 0.72 | 58.3 | 0.80 | 0.86 | 6.2 |
| ResUnet++ | 0.82 | 0.71 | 51.6 | 0.80 | 0.85 | 5.9 |
| **ROIsGAN (Ours)** | **0.85** | **0.75** | **31.6** | **0.84** | **0.87** | **4.5** |

### B. Cohort 1-NeuN Dataset

For NeuN-stained images (Table 2), which emphasize neuronal nuclei, ROIsGAN consistently leads across performance metrics, achieving a mean Dice coefficient of 0.79. This represents a modest yet notable improvement over ResU-Net++ (0.78) and a more substantial gain over U-Net, U-Net++, and Attention U-Net (0.77).

ROIsGAN's performance is particularly pronounced in Recall (0.82) and ASSD (6.7), where it outperformed ResU-Net++ (Recall: 0.81, ASSD: 7.3) and all other baselines. This suggests that ROIsGAN's region-guided hybrid loss function effectively mitigates over-segmentation errors, enhancing both sensitivity and structural accuracy in NeuN-stained ROI segmentation. The IoU (0.68) remains comparable to the best-performing baseline, while the HD (43.5) marginally lags behind ResU-Net++ (41.2). Nonetheless, ROIsGAN maintains high fidelity in capturing the intricate geometry of neuronal structures, thereby confirming its robust generalizability to morphologically rich IHC signals (Figure 3. c).

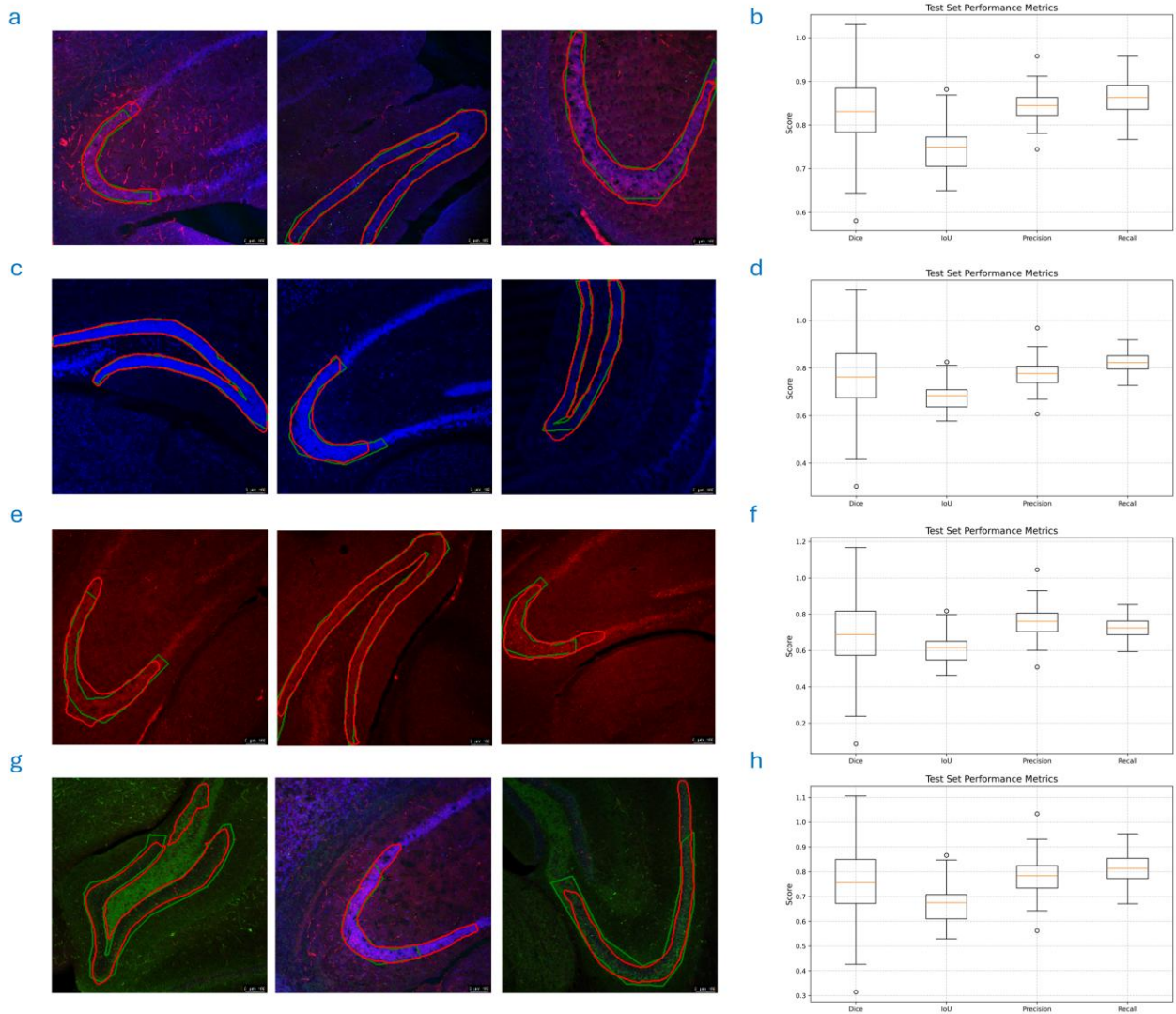

*Figure 3. ROIsGAN models segmentation performance across modalities: green contours represent ground truth and red contours depict the predicted mask: a) prediction on cohort 1 multiplexed dataset, b) performance on c1 multiplexed dataset, c) NeuN dataset d) NeuN dataset performance, e) cFos dataset, f) cFos dataset performance, g) extended multiplexed samples, h) extended multiplex dataset.*

*Table 2. Performance comparison with existing state-of-the-art model on NeuN stained images*

| Model | Dice | IoU | HD | Precision | Recall | ASSD |
|---|---|---|---|---|---|---|
| Unet | 0.77 | 0.65 | 60.5 | 0.77 | 0.78 | 8.5 |
| Unet++ | 0.77 | 0.66 | 55.6 | 0.77 | 0.79 | 7.7 |
| AttentionUnet | 0.77 | 0.66 | 54.7 | 0.75 | 0.81 | 7.8 |
| ResUnet++ | 0.78 | 0.68 | 41.2 | **0.78** | 0.81 | 7.3 |
| **ROIsGAN (Ours)** | **0.79** | **0.68** | 43.5 | 0.77 | **0.82** | **6.7** |

### C. Cohort 1-cFos Dataset

The cFos-stained dataset constitutes the most challenging segmentation task, attributable to its sparse distribution and low signal-to-noise ratio inherent in the staining pattern. Despite these challenges, ROIsGAN attains a Dice coefficient of 0.72, outperforming ResU-Net++ (0.70) and substantially surpassing U-Net (0.63), U-Net++ (0.64), and Attention U-Net (0.62), as shown in Table 3.

*Table 3. Performance comparison with existing state-of-the-art model on cFos stained images*

| Model | Dice | IoU | HD | Precision | Recall | ASSD |
|---|---|---|---|---|---|---|
| Unet | 0.63 | 0.51 | 68.8 | 0.67 | 0.65 | 13.0 |
| Unet++ | 0.64 | 0.51 | 62.9 | 0.70 | 0.63 | 13.2 |
| AttentionUnet | 0.62 | 0.50 | 63.7 | 0.71 | 0.60 | 14.4 |
| ResUnet++ | 0.70 | 0.58 | 63.4 | 0.71 | **0.73** | **11.2** |
| **ROIsGAN (Ours)** | **0.72** | **0.61** | 49.5 | **0.75** | 0.72 | 11.8 |

In addition to Dice, ROIsGAN achieves superior IoU (0.61), Precision (0.75), and Recall (0.72), reflecting enhanced performance in segmenting sparse, and morphologically irregular ROIs. Although its ASSD (11.8) slightly trails that of ResU-Net++ (11.2), the notably lower HD (49.5) compared to ResU-Net++ (63.4) underscores ROIsGAN's superior boundary localization and reduced susceptibility to spurious edge predictions (Figure 3. a).

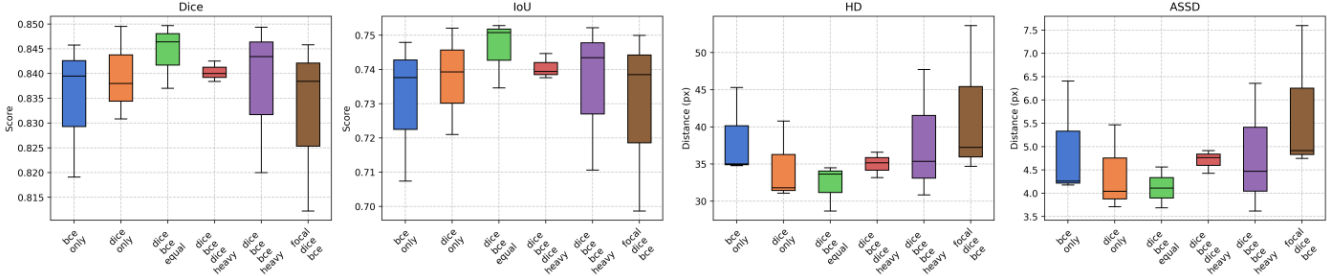

*Figure 4. Ablation study evaluating discriminator loss functions: BCE-only, Dice-only, Dice+BCE with various weightings, and Focal+BCE. The proposed Dice+BCE (equal weighting) formulation achieves the most consistent and robust performance across all segmentation metrics.*

### D. Extended Multiplexed Dataset

On the larger and more diverse extended multiplexed dataset ROIsGAN outperforms all baseline models across all evaluation metrics, achieving a Dice coefficient of 0.79, the highest among all methods (Table 4).

*Table 4. Performance comparison with existing state-of-the-art model on multi-cohort Extended Multiplexed image dataset*

| Model | Dice | IoU | HD | Precision | Recall | ASSD |
|---|---|---|---|---|---|---|
| Unet | 0.76 | 0.65 | 65.0 | 0.74 | 0.82 | 10.3 |
| Unet++ | 0.76 | 0.63 | 69.0 | 0.74 | 0.80 | 10.0 |
| AttentionUnet | 0.75 | 0.63 | 74.2 | 0.73 | 0.82 | 11.3 |
| ResUnet++ | 0.78 | 0.67 | 57.1 | 0.78 | 0.81 | 9.3 |
| **ROIsGAN (Ours)** | **0.79** | **0.68** | **55.6** | **0.79** | **0.81** | **8.5** |

This represents a clear improvement over traditional architectures, including U-Net, U-Net++, Attention U-Net, and ResU-Net++. ROIsGAN also achieves the best IoU (0.68), Precision (0.79), Recall (0.81), and the lowest boundary error metrics HD (55.6) and ASSD (8.5), highlighting its superior segmentation accuracy and structural fidelity. These results reflect the effectiveness of ROIsGAN's region-guided adversarial training in promoting anatomically precise predictions.

The modest Dice drop from 0.85 on Cohort 1-multiplexed to 0.79 on the extended-multiplexed dataset reflects increased variability in staining and anatomical structure. Nevertheless, ROIsGAN maintains strong performance, demonstrating its adaptability and generalizability across heterogeneous imaging conditions.

### E. Ablation Study

To investigate the influence of discriminator supervision on overall segmentation performance, we conducted an ablation study comparing various loss configurations: BCE-only, Dice-only, Dice+BCE with equal and imbalanced weightings, and a Focal+BCE hybrid. These loss variants were used exclusively to train the discriminator, while the generator loss remained fixed. Figure 4 presents the resulting segmentation performance across Dice, IoU, Precision, and Recall metrics.

Among all variants, the equal-weighted Dice+BCE configuration consistently yielded the best overall performance. It achieved high median Dice and IoU scores with low variance while achieving low HD and ASSD values. Dice-heavy configurations yielded higher HD and ASSD due to boundary inaccuracies, whereas BCE-only and Focal+BCE losses exhibited increased variability across all metrics. These results suggest that training the discriminator with both region- and boundary-level feedback encourages more stable and effective generator updates. Consequently, we adopt the Dice+BCE (equal) loss for the discriminator in our final ROIsGAN framework.

In addition, we investigated the effectiveness of our postprocessing pipeline across all datasets. While region-based metrics such as Dice, IoU, precision, and recall remained stable, substantial improvements were observed in surface-based metrics (HD and ASSD). In preliminary trials across multiple hippocampal datasets, post-processing yielded marginal changes in region-based metrics (Dice improvement ≤ 0.01). However, it led to substantial reductions in boundary error, decreasing HD by 65–80% and Average Symmetric Surface Distance by 84–94%, highlighting its efficacy in refining anatomical boundaries (Table 5). This indicates that post-processing predominantly refined the segmentation boundaries and removed outliers without significantly altering the overall segmentation region.

*Table 5. Ablation study evaluating the impact of post-processing across different datasets.*

| Dataset | Post-processing | Dice | IoU | HD | Precision | Recall | ASSD |
|---|---|---|---|---|---|---|---|
| Multiplexed | No | 0.84 | 0.74 | 160.2 | **0.85** | 0.85 | 70.7 |
| | Yes | **0.85** | **0.75** | **31.6** | 0.84 | **0.87** | **4.5** |
| NeuN | No | 0.79 | 0.68 | 161.5 | 0.77 | 0.82 | 73.0 |
| | Yes | **0.79** | **0.68** | **43.5** | 0.77 | **0.82** | **6.7** |
| cFos | No | 0.72 | 0.60 | 168.7 | 0.75 | 0.72 | 74.7 |
| | Yes | **0.72** | **0.61** | **49.5** | 0.75 | 0.72 | **11.8** |
| Extended multiplexed | No | 0.78 | 0.67 | 157.9 | 0.79 | 0.81 | 70.0 |
| | Yes | **0.79** | **0.68** | **55.6** | 0.79 | **0.81** | **8.5** |

### F. Explainability via Grad-CAM

To assess the anatomical fidelity of ROIsGAN's predictions, we generated Grad CAM heatmaps over representative test samples from all four datasets (Figure 5). In each case, the Grad CAM map highlights the image regions that most influenced the generator's segmentation of hippocampal subregions. Across staining modalities, activation predominantly coincided with true hippocampal contours. In both cFos and NeuN cohorts, high-activation regions closely track subregion boundaries, reflecting the model's reliance on well-defined structural features. In the Extended Multiplexed dataset, Grad CAM activations remained sharply localized around the hippocampal region, even in the presence of complex GAD67 staining, suggesting robustness to staining variability.

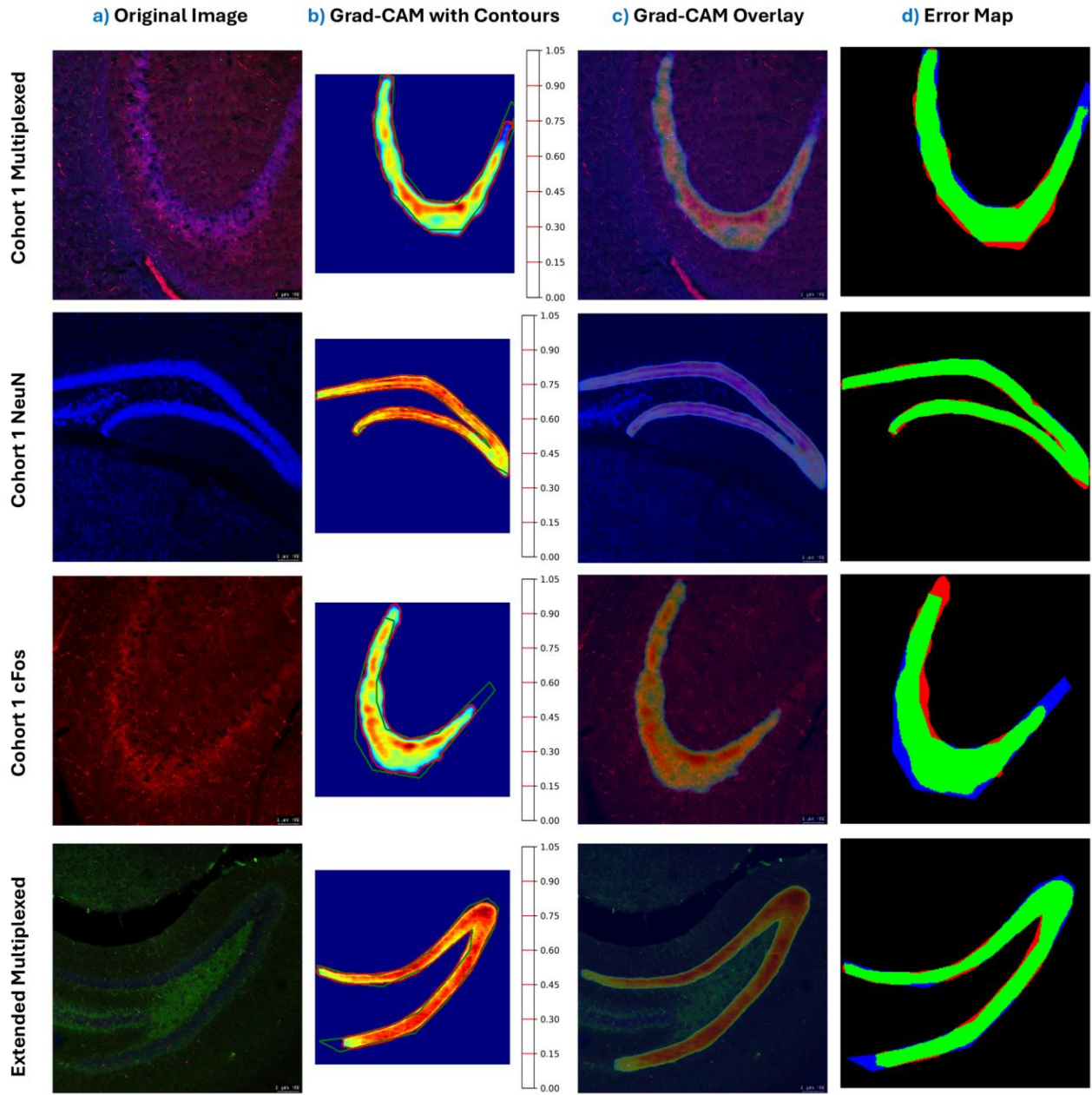

*Figure 5. Explainable AI results for hippocampus segmentation across four datasets. Each row represents one sample: (a) Original image, (b) Grad-CAM heatmap (jet colormap) with true (green) and predicted (red) contours, (c) Grad-CAM overlay (jet colormap, 30% transparency) on original image, (d) Error map (true positives in green, false positives in red, false negatives in blue). Visualizations highlight regions driving segmentation decisions, with error maps indicating performance variations.*

Figure 5.b overlays the Grad CAM map with predicted (green) and ground truth (red) contours, confirming strong spatial correspondence. Grad CAM overlays in Figure 5.c emphasize that attention is confined to anatomically relevant regions. The error maps in Figure 5.d, color-coded for true positives (green), false positives (red), and false negatives

(blue), show that most segmentation discrepancies occur near subregion boundaries, where staining ambiguity is greatest.

These qualitative findings validate ROIsGAN's structure-aware learning. However, Grad CAM's coarse resolution may limit the interpretation of fine features, and future work could explore higher-resolution attribution methods to further elucidate model decision processes.

## IV. DISCUSSION

We introduce the first publicly available immunohistochemistry hippocampal segmentation dataset, comprising four datasets with a total of 1,558 murine hippocampal images stained with combinations of cFos, NeuN, ΔFosB, and GAD67, including single-stain and multiplexed modalities. This comprehensive dataset captures structural, activity-related, and plasticity-associated signals, addressing a critical gap in histological neuroimaging by enabling automated segmentation of DG, CA1, and CA3 subregions. It provides a robust benchmark for evaluating segmentation algorithms under real-world staining and anatomical variability, facilitating downstream analyses in neuroscience, such as quantitative studies of neuronal activity and disease phenotyping, and paving the way for advanced segmentation tools.

Leveraging this comprehensive dataset, we propose ROIsGAN, a novel region-guided GAN framework tailored for hippocampal subregion segmentation. ROIsGAN achieves superior performance across all four datasets, improving Dice scores by 0.01–0.10 over established segmentation baselines (Tables 1–4). It consistently yields the lowest ASSD across all datasets and reduces HD in multiplexed and extended multiplexed datasets. While ROIsGAN exhibits a marginally higher HD on NeuN-stained images compared to ResU-Net++ (43.5 vs. 41.2 pixels), it demonstrates enhanced boundary delineation consistency, as evidenced by qualitative assessments (Figure 3). Notably, ROIsGAN maintains robust segmentation accuracy on cFos-stained images, despite their low signal-to-noise ratio and sparse labeling, underscoring its resilience to challenging imaging conditions. The region-guided discriminator, a key innovation, integrates Dice and binary cross-entropy losses to produce spatial realism maps, enforcing subregion-level structural coherence, particularly at ambiguous boundaries. This contrasts with conventional GAN discriminators, which operate as image-level classifiers and lack fine-grained spatial guidance. Ablation studies on the loss functions demonstrate the effectiveness of our proposed region-guiding hybrid loss in the hippocampal subregion segmentation over the traditional approaches (Figure 4).

Despite the superior performance of ROIsGAN in all modalities, we observed a drop in Dice from Cohort 1 (0.85) to the extended multiplexed dataset (0.79) attributed to inter-cohort variation in staining protocols and hippocampal morphology. This limitation underscores the challenge of generalizing across heterogeneous datasets. Additionally, the model's reliance on high-quality annotations, compounded by potential human labeling and instrumental errors during IHC and imaging, may limit its applicability to noisy or incomplete datasets. To address these limitations, future research will explore domain-adaptation, stain-normalization, semi-supervised learning with weakly labeled data, and transfer learning to enhance robustness and scalability across diverse imaging protocols, thereby broadening ROIsGAN's utility in neuroimaging.

From a translational perspective, ROIsGAN significantly reduces the manual annotation effort required for hippocampal subregion segmentation. Manual delineation of densely stained hippocampal subfields is labor-intensive, time-consuming, and susceptible to inter-observer variability, particularly in large-scale studies of activity-dependent plasticity and disease phenotyping. ROIsGAN automates this intricate process, providing neuroscientists with a rapid, reproducible, and robust tool. It holds enormous potential for streamlining downstream analyses of tissue images, such as identifying group differences in cFos expression—a widely utilized marker for neuronal activity—particularly in brain regions with thousands of immunoreactive neurons. Additionally, automated segmentation mitigates the researcher-dependent variability associated with manual ROI selection, thus enhancing the efficiency, consistency, and reproducibility of neuroscientific research.

This region-guided adversarial framework establishes a foundational basis for automated histopathology in neuroimaging. Its adaptability to other brain regions and tissue types promises to significantly advance the scalability and precision of medical imaging analyses, facilitating broader applications across neuroscience and clinical diagnostics.

## V. CONCLUSION

We present ROIsGAN, a region-guided generative adversarial network, alongside four novel and comprehensive murine hippocampal IHC datasets tailored for automated segmentation of DG, CA1, and CA3 subregions from fluorescent tissue images. These datasets uniquely capture structural, neuronal activity-related, and plasticity-associated signals, addressing a critical gap in neuroimaging benchmarks. ROIsGAN integrates a hybrid Dice–binary cross-entropy loss within an adversarial training framework, guiding the generator toward significantly enhanced segmentation precision, particularly at anatomically ambiguous boundaries. Our extensive evaluations demonstrate Dice scores ranging from 0.72 to 0.85, consistently outperforming established methods, including U-Net, U-Net++, Attention U-Net, and ResU-Net++ across multiple staining modalities. By providing these datasets and an effective segmentation tool, we establish a foundational resource for enabling, scalable, and precise hippocampal tissue image analyses in neuroscience research. Future work will focus on cohort-specific data augmentation, enhanced generalization strategies, and integration with automated phenotyping pipelines. Our datasets, segmentation model, and corresponding source code are publicly accessible at: https://github.com/MehediAzim/ROIsGAN .